\documentclass{cup-hpl}

\usepackage{soul}
\usepackage{xcolor}
\usepackage{booktabs}
\usepackage[separate-uncertainty = true]{siunitx}

\begin{document}

\shorttitle{A multi-shot target-wheel assembly for high-repetition-rate, laser-driven proton acceleration}
\shortauthor{J. Pe\~nas, A. Bembibre, et al.}

\title{A multi-shot target-wheel assembly for high-repetition-rate, laser-driven proton acceleration}

\author[1,*]{J. Pe\~nas}
\author[1,*]{A. Bembibre}
\author[1]{D. Cortina-Gil}
\author[2]{L. Mart\'in}
\author[1]{A. Reija}
\author[3]{C. Ruiz}
\author[4]{M. Seimetz}
\author[1]{A. Alejo$^{\dagger,}$}
\author[1]{J. Benlliure$^{\ddagger,}$}\corresp{$^\dagger$aaron.alejo@usc.es; $^\ddagger$j.benlliure@usc.es}

\address[1]{Instituto Galego de Física de Altas Enerxías, Universidade de Santiago de Compostela, Santiago de Compostela, Spain}
\address[2]{Laboratorio Láser de Aceleración y Aplicaciones, RIAIDT, Universidade de Santiago de Compostela, Santiago de Compostela, Spain}
\address[3]{Instituto Universitario de F\'isica Fundamental y Matem\'aticas y Dpto. de Did\'actica de la Matem\'atica y de las Ciencias Experimentales, Universidad de Salamanca, Salamanca, Spain}
\address[4]{Instituto de Instrumentaci\'on para Imagen Molecular (I3M), CSIC-Universitat Polit\`ecnica de Val\`encia, Valencia, Spain}
\address[*]{These authors contributed equally to this work.}

\begin{abstract}
A multi-shot target assembly and automatic alignment procedure for laser-plasma proton acceleration at high-repetition-rate are introduced. The assembly is based on a multi-target rotating wheel capable of hosting $>$5000 targets, mounted on a three-dimensional motorised stage to allow rapid replenishment and alignment of the target material between laser irradiations. The automatic alignment procedure consists of a detailed mapping of the impact positions at the target surface prior to the irradiation that ensures stable operation of the target, which alongside the purpose-built design of the target wheel, enable the operation at rates up to 10\,Hz. Stable and continuous laser-driven proton acceleration is demonstrated, with observed cut-off energy stability about 15\%.
\end{abstract}


\maketitle

\section{Introduction}

The acceleration of ions from the interaction of a high-power laser with a plasma has attracted a growing interest over the last two decades~\cite{daido2012review, macchi2013ion}. Efficient acceleration of light ions can be achieved through a variety of established and emerging accelerating mechanisms, whose dominance depending on the laser and target characteristics. Arguably, the so-called Target Normal Sheath Acceleration (TNSA) mechanism is the most-established and robust route to accelerate light ions to multi-MeV energies~\cite{robson2007scaling, fuchs2006laser}. In TNSA, the laser interacts with an overdense plasma, typically generated by the pedestal preceding the main pulse in high power systems. During the interaction, a large number of fast electrons from the front surface are driven into the target. These fast electrons will reach the target rear surface, where a fraction will leave the target and generate a quasi-electrostatic, capacitor-like field at the rear surface, with values on the order of TV\,m$^{-1}$. Such strong electric field will ionise the atoms on the rear surface and accelerate the ions to multi-MeV energies. The TNSA-driven ion beams are characterised by their multi-species nature, but dominated by protons, with quasi-exponential spectra extending up to a sharp cut-off energy, and emitted with a divergence of up to tens of degrees.

Parallel to the studies of laser-driven ion acceleration, there has been a significant progress in laser technology, not only towards achieving increasingly larger powers, but also towards the development of multi-TW laser systems with increasingly higher repetition rates. The advent of these multi-Hertz high-power laser systems is leading to a growing need for novel target systems capable of operating at such rates. Due to the destruction of the target following the interaction with the laser, an appropriate system needs to be capable of replenishing the target and position it on the focal plane of the laser with micron-level precision, as given by the Rayleigh length for the short $f$-number optics used in this type of experiments. Furthermore, a future laser-based ion accelerator will be required to operate continuously at these rates for extended periods of time, and therefore a suitable target system must be able to host thousands of targets.

Several alternatives are being actively studied as potential target systems~\cite{bembibre2023long}. Some promising recent developments include the use of liquid targets~\cite{morrison2018mev}, liquid crystal targets \cite{poole2014liquid, schumacher2017liquid}, high-density gas jets~\cite{palmer2011monoenergetic, puyuelo2019laser}, or cryogenic solid hydrogen targets~\cite{tebartz2017creation, polz2019efficient}, all of which would ensure the operation for extended periods of time. However, these solutions still face major challenges, such as the shape and profile manipulation, micrometric positioning, and restrictions in operation due to the high-vacuum level required by the laser systems. For these reasons, target systems based on the replenishment of foil-based solid targets remain as the most common solution, typically on the form of tape-drive systems or multi-target holder systems.

Tape-drive-based targets allow for tens of thousands of shots, that at 10 Hz would correspond to almost one hour of continuous operation~\cite{nayuki2003thin, noaman2017statistical, condamine2021high, xu2023versatile}. The main drawback of these systems is the limited variety of tape materials and thicknesses suitable to withstand the mechanical stress caused by the continuous movement. In this context,  multi-target holder systems appear as an appealing alternative, thanks to the flexibility of using a rigid structure to support the target foils, allowing for a broad variety of suitable target materials and thicknesses~\cite{gao2017automated}. However, these configurations present two major limitations, namely the relatively reduced number of shooting positions, typically hosting fewer than 1000 targets; and the reduced repetition rate at which they can be operated, due to the need to replace and realign with micron-level precision after each irradiation. Recent developments have tried to tackle these limitations in order to extend the usability of multi-target holder systems.

In order to increase the number of targets, Gao \etal~\cite{gao2017automated} proposed a system based on a metallic target wheel hosting target plates accommodating up to $\sim1700$ impact positions. However, the maximum operation rate of this target assembly is limited to \SI{0.5}{Hz}, as given by the relaxation time to reduce the vibrations of the target wheel after the movement of the motorised stages. Another appealing alternative to increase the shooting positions is the use of MEMS technology to fabricate micro- and nano-targets on silicon wafers~\cite{spindloe2016high, zaffino2018preparation, gershuni2019gatling}. This technology allows for manufacturing large volumes of micro-targets made of different materials and thicknesses, including composites of various distinct layers. For instance, Gershuni \etal~\cite{gershuni2019gatling} propose a target delivery system based on \SI{100}{mm}-diameter Si wafers where hundreds of microtargets can be created using MEMS technology. However, the online measurement and closed-loop correction sequence for target realignment limited the repetition rate to \SI{0.2}{Hz}.


Here we report on a multi-shot target assembly based on a rotating wheel capable of hosting $>$5000 targets, and compatible with operation at a repetition rate of \SI{10}{Hz}. Furthermore, we describe the procedure implemented to ensure automatic shot-to-shot replenishment and realignment of each target at \SI{10}{Hz}, based on a few-minute measurement for the pre-characterisation of the shooting positions with a high-precision industrial sensor, that allows for the positioning of the targets on the focal plane with a precision of $\sigma=\SI{3.5}{\micro\m}$. Experimental results on laser-driven ion acceleration from the developed target and alignment method at the \textit{Laser Laboratory for Acceleration and Applications} (L2A2)~\cite{benlliure2019validation, alejo2022characterisation} are presented, demonstrating the stable, continuous operation for $>$1000 shots with deviations in the cut-off energy of the measured proton spectra of $\sim15\%$, limited by the stability of the laser system. The rest of the paper is structured as follows. The target assembly, including a detailed description of the design requirements for the target wheel depending on the stages and minimum repetition rate, is included in Sec.~\ref{sec:assembly}. The target pre-mapping and automatic positioning procedure are described in Sec.~\ref{sec:mapping}. Finally, the experimental setup and results on ion acceleration using the rotating wheel target are discussed in Sec.~\ref{sec:acceleration}.

\section{The multi-shot target assembly}\label{sec:assembly}
\begin{figure}[h]
  \centering
  \includegraphics[width=\linewidth]{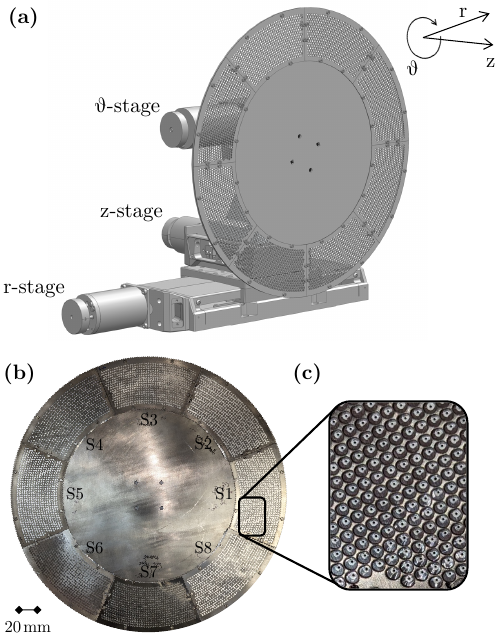}
  \caption{\textbf{(a)} Drawing of the target assembly, depicting the three motorised stages and a rotating wheel. \textbf{(b)} Picture of a target wheel design for \SI{10}{Hz} operation. \textbf{(c)} Zoomed-in picture of the target wheel, showing the craters on the targets after their irradiation.}
  \label{Fig:WheelDesign}
\end{figure}

The target assembly developed is based on a multi-shot rotating wheel attached to a three-dimensional motorised rig to ensure the shot-to-shot replacement and positioning of the target material (Fig.~\ref{Fig:WheelDesign}(a)). The motorised rig consists of a rotational stage (PimiCos T-65N) that enables the rotation of the target wheel around its axis ($\vartheta$); a linear stage (PimiCos DT-65N) that enables the translation along the target normal, or \textit{longitudinal} direction ($z$); and a linear stage (PimiCos LS-110) that enables the translation along the direction perpendicular to the target normal, or \textit{radial} direction ($r$). The combined motion of the rotational and radial stages allows to change the irradiated target between laser shots, whereas a combined motion of the radial and longitudinal stages allows to displace the wheel along the laser focal direction and position each target at the focal plane. The spatial resolution of the rotational, longitudinal, and radial stages is \SI{20}{\micro deg}, \SI{10}{nm}, and \SI{20}{nm}, respectively, significantly lower than the required tolerance for the positioning of the target at the focal plane, typically on the order of \SI{10}{\micro\m} (see Sec.~\ref{sec:mapping}).

The key element of the target assembly is the multi-shot rotating wheel. In our case, the wheel is formed by a base \SI{304}{\milli\m}-diameter, \SI{2}{\milli\m}-thick aluminium disk that can be directly attached to the rotational stage, and capable of hosting up to 8 target sectors. Each sector consists of a target foil of choice sandwiched between two planar plates of \SI{200}{\micro\m} in thickness, with pre-drilled holes that allow the irradiation of the target by the laser (Fig.~\ref{Fig:WheelDesign}(b)). These planar plates allow to ensure the stability of the foil throughout the operation, avoiding the deformation of the foil in impact positions adjacent to the irradiated targets (Fig.~\ref{Fig:WheelDesign}(c)), which could otherwise result in changes greater than \SI{100}{\micro\m}. The actual distribution, which sets the maximum repetition rate and number of impact positions of the wheel, is limited by factors such as the maximum speed and range of motion of the stages, as well as the minimum size of and distance between the holes on the sector plates. In order to maximise the number of irradiation points, the pre-drilled holes are distributed along concentric circular patterns, where the diameter of each hole is kept at \SI{2}{\milli\m} to prevent damage by the laser. Given the thinness of the plates, a minimum center-to-center distance between holes of $d_{H}=\SI{2.5}{\milli\m}$ was required in order to ensure mechanical stability.

Considering the aforementioned constraints, the maximum repetition rate of operation of the target for a row of the circular pattern at radius $R_H$ is limited by the maximum velocities for the rotational ($\dot{\vartheta}_{\max}=\SI{22}{\degree\per\s}$) and radial ($\dot{r}_{\max} = \SI{25}{\milli\m\per\s}$) stages. In our case, a key requirement for the wheel is its capability to operate at \SI{10}{Hz}, resulting in a maximum target-replacement time of $\tau_R=$\SI{100}{ms}. Therefore, the impact positions must verify that $d_{H} \leq \tau_R/\dot{r}_{\max} = \SI{2.5}{mm}$, and $d_H\leq R_H\,\dot{\vartheta}_{\max}\,\tau_R = 0.038 R_H$. In addition to setting a maximum center-to-center distance between holes, already fulfilled by our wheel design, these conditions establish a minimum radius for the circular pattern of holes, corresponding to $R_H\geq \SI{65}{mm}$ for our conditions. However, it should be noted that larger radii will result in a greater number of impact positions. For this reason, the inner radius in our case was designed to be $\SI{100}{mm}$, whereas the number of concentric arcs was limited to 18, as given by the center-to-center inter-hole distance and the range of motion of the radial stage (\SI{50}{mm}). As a result, each sector can host up to $650$ targets, leading to a total of $5200$ targets for the entire wheel. Such a large number of shooting positions not only represents a significant increase with respect to similar systems based on multi-target holders~\cite{gao2017automated, chagovets2021automation}, but also makes this design competitive with respect to other solutions such as tape-drives or MEMS targets, particularly considering that it can be further increased through careful choices of the inner and outer radii of the shooting positions. 

\section{Target automatic positioning system}\label{sec:mapping}
The tolerance for the target positioning is defined by the length in which the laser beam remains focussed, given by its Rayleigh length, typically on the order of $\sim$\SI{10}{\micro\m} for the focussing optics employed in laser-driven ion acceleration experiments. This accuracy is beyond the intrinsic precision of the target system, limited by factors such as mechanical deformations of the wheel and target foil, or the wobbling associated to the rotational stage. Therefore, some form of positioning system is required in order to ensure the placement of each target on the focal plane without the need for individual alignment. 

\begin{figure}[b!]
  \includegraphics[width=\linewidth]{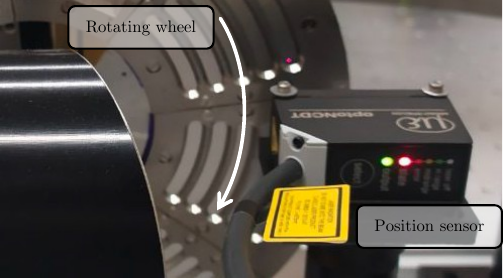}
  \centering
  \caption{Picture of the setup used for the validation of the target positioning system.}
  \label{Fig:Sensor}
\end{figure}

In our case, we have implemented an automatic correction system based on adjusting the target position according to a detailed 3D mapping of the impact positions of the laser pulses on target, allowing for operation at higher repetition rates than other methods based on online, live correction procedures. To obtain this pre-map, the coordinates of the desired impact positions on the target wheel are defined by its mechanical design, while the longitudinal coordinate, or displacement along the target normal, is measured with an OptoNCDT ILD1320 position sensor by Micro-Epsilon, with a reproducibility of \SI{1}{\micro\m} (Fig.~\ref{Fig:Sensor}). 
It should be noted that the sensor cannot operate in vacuum, and therefore the entire wheel characterisation must be performed in atmospheric conditions prior to the irradiation. In order to minimise the time required to measure the pre-map, this process is typically performed at the maximum stage speed, resulting in an entire wheel being evaluated in a few minutes, at a rate of $\sim10$ points per second. Our method represents a significant improvement with respect to conventional target characterisation techniques, e.g. \SI{100}{min} required to characterise 1000 impact positions by Chagovets \etal~\cite{chagovets2021automation}.

\begin{figure}[t]
  \includegraphics[width=0.48\textwidth]{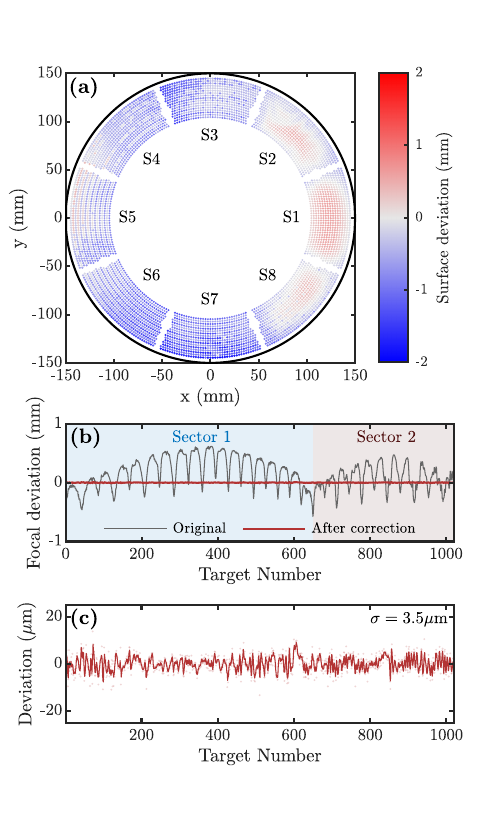}
  \caption{\textbf{(a)} 3D surface map of the aluminium target foils installed at the multi-target wheel. \textbf{(b)} Surface profile of the first $\sim$1000 impact positions before (black curve) and after (red curve) correction. \textbf{(c)} Zoomed-in view of the deviation of the impact positions after correction shown in (b). Each impact position appears represented by an individual marker, and the straight line shows the 3-point moving average of the deviations.}
  \label{Fig:Map}
\end{figure}

A pre-map measured using the distance sensor for the wheel in Fig.~\ref{Fig:WheelDesign}(b) is shown in Fig.~\ref{Fig:Map}(a). Large deviations can be observed between the different shooting positions, with variations greater than \SI{1}{\milli\m} between different regions of the wheel. Furthermore, these deviations are measured not only between different sectors, but also between consecutive targets within the same sector, indicating a deformation of the target surface probably caused by the procedure to sandwich the target foil between the pre-drilled plates, as pointed out by the greater deviations in the regions away from the edges where the target foil is clamped.

The information contained in the pre-map can be subsequently used to automatically correct the position of each target. In our case, a control software was developed using LabVIEW to handle the motion of the three motorised stages in the target assembly. In a first step, the system calculates the initial deviation with respect to the desired plane for each impact position, as shown in Fig.~\ref{Fig:Map}(b) [black line]. This information is used to calculate the required combined motion of the stages in order to place the target at the desired plane while ensuring that the impact position remains unchanged. 

To validate this technique, a position verification was performed, based on the measurement by the distance sensor of the displacement of the impact positions while the correction was being applied. This process was performed in the same conditions as the real laser irradiations, including equivalent motion profile for the different stages and operation at a repetition rate of \SI{10}{Hz}. The results of this procedure for the first $\sim$1000 impact positions are shown in Fig.~\ref{Fig:Map}(b) [red], which clearly indicate an improvement with respect to the original movements without correction. For clarity, the same results are shown with a different scale in Fig.~\ref{Fig:Map}(c), where the individual measurements by the sensor are depicted by the markers, and the straight line represents a 3-point moving average of the experimental data. The distribution of measurements shows a standard deviation of $\sigma=\SI{3.5}{\micro\m}$, well below the Rayleigh length of the laser system.

It should be noted that, albeit the developed control system has been shown to be capable of accurately positioning each target at the desired plane, it is crucial to ensure that such plane correspond to the focal plane of the laser beam. In our case, this information is fed into the control system by manually bringing one of the targets to the laser focal plane. Different techniques are available for the alignment of solid targets, such as the speckle technique~\cite{alexeev2017determination}, the retro-imaging technique~\cite{carroll2011assessment, kumar2019alignment}, or the direct imaging technique~\cite{singh2016diagnostic}. For the results here presented, the latter was used, in which the focal plane of the laser is initially found with a high-magnification imaging system, allowing to place a back-illuminated target at the same plane by ensuring the same optical system is imaging the rear surface of the target.

\section{Proton acceleration at L2A2}\label{sec:acceleration}
\subsection{Experimental Setup}\label{sec:setup}
\begin{figure}[b]
  \includegraphics[width=\linewidth]{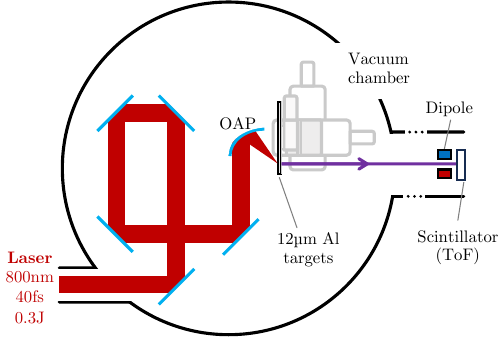}
   \caption{Schematic representation of the experimental set-up used at L2A2.}
  \label{Fig:Setup}
\end{figure}

Laser-driven proton acceleration using the developed rotating wheel target and automatic alignment procedure has been studied experimentally using the setup schematically depicted in Fig.~\ref{Fig:Setup}. The experiments were performed utilising the STELA laser system, hosted at the Laser Laboratory for Acceleration and Applications (L2A2, Universidade de Santiago de Compostela)~\cite{benlliure2019validation, alejo2022characterisation}, which provided p-polarised, \SI{800}{nm}-wavelength pulses at a repetition rate of \SI{10}{Hz}, containing energies of up to \SI{0.3}{J} on target, and compressed to a duration of $\sim$\SI{40}{fs}.

Inside a vacuum chamber maintained at a pressure of $\sim$\SI{1e-6}{mbar}, the laser beam was focussed onto the targets with a \SI{45}{\degree} off-axis parabolic mirror (\textit{f}/2.8) down to a $\sim$\SI{5}{\micro\m} focal spot size (Full-Width-at-Half-Maximum), reaching intensities of $\sim$\SI{3e19}{\W\per\cm\squared}. Aluminium foils of \SI{12}{\micro\m} thickness were mounted on the rotating target wheel and used as targets. The aforementioned procedure was employed for the replacement and positioning of the targets on the focal plane. The automatic control system was synchronised with the laser through a common trigger in order to ensure that each movement was completed between the irradiations. 

The laser-driven proton beam was characterised using the time-of-flight (ToF) technique, based on the temporally-resolved measurement of the signal produced by the protons reaching a detector placed at a distance of \SI{2}{m} from the interaction point. In order to prevent the detection of electrons, a dipole magnet was placed along the path to deviate electrons while allowing ions and photons into the detector. The detector consisted of a fast plastic scintillator (NE102A) covering an area of $25\times$\SI{25}{\mm\squared} ($\sim$\SI{0.16}{\milli\steradian}), with a thickness of \SI{5}{mm}, sufficient to stop incoming protons with energies up to \SI{22.5}{MeV}. In order to reduce the external light noise in the detection system, the scintillator piece was covered by a \SI{4}{\micro\m}-thick layer of aluminised Mylar. The signal emitted by the scintillator was collected and detected using a photo-multiplier tube (PMT). In order to minimise the electromagnetic noise caused by the laser-plasma interaction, the PMT was shielded and placed away from the radiation by means of optical fibres for its coupling with the scintillator. Due to the large optical signals generated in the scintillator, a neutral density filter OD1 was placed between the optical fibres and the PMT~\cite{seimetz2015calibration}.

\subsection{Experimental results}\label{sec:results}
\begin{figure}[tbh!]
  \includegraphics[width=\linewidth]{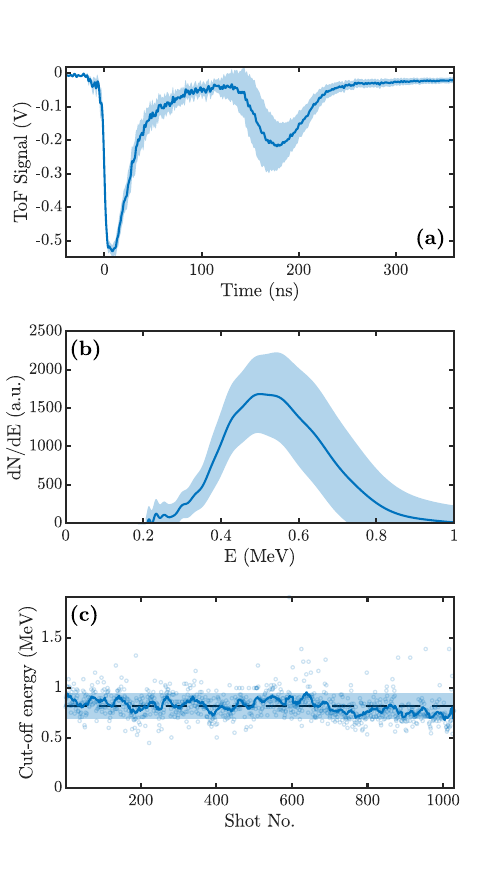}
   \caption{\textbf{(a)} Time-of-flight signal of the laser-driven ions. The line and shaded area represent, respectively, the average and standard deviation of the signal detected for 1032 consecutive shots. \textbf{(b)} Energy spectrum obtained for the data in (a). \textbf{(c)} Evolution of the proton cut-off energy, where each individual marker indicates the peak energy for each irradiation, the dark line shows the 20-period moving average, and the shaded area shows the standard deviation of the cut-off energies around the \SI{0.81}{MeV} mean energy.}
  \label{Fig:Spectra}
\end{figure}

The time-of-flight signal measured for 1032 consecutive shots, corresponding to the mapped target points in Fig.~\ref{Fig:Map}(b-c), is shown in Fig.~\ref{Fig:Spectra}(a), where the line and shaded area represent the average and the standard deviation of the signals, respectively. Two distinct peaks can be identified on the signal. A first peak corresponds to the so-called gamma-flash, produced by high-energy photons generated during the laser-plasma interaction, which can be used as a reference time for the arrival of the laser pulses. The second peak in the signal corresponds to the incoming charged particles, protons and heavier ions, reaching the scintillator at a later time depending on their energy $E_i$. 

The energy spectrum of the ion beam can be reconstructed from the ToF signal. Considering the non-relativistic limit, the detection time and energy can be related through the expression,
\begin{equation}
	E_i = \frac{1}{2} m_i \left(\frac{D}{\Delta t_i + D/c}\right)^2,
\end{equation}
where $m_i$ is the ion mass, $D$ is the length of the flight path, $c$ is the speed of light, and $\Delta t_i$ is the delay of the signal with respect to the arrival of the gamma flash. It should be noted that, as previously discussed, laser-driven ion beams are characterised for having a multi-species nature, but are heavily dominated by protons from the contaminant layers on the target surface, and to a lesser extent by carbon and oxygen ions. Unlike other diagnostic tools, such as Thomson parabola spectrometers, ToF-based diagnostics cannot discriminate between different ion species. However, given the dominance of the protons within the beam, as well as the presence of the aluminised mylar capable of fully stopping carbon and oxygen ions with energy up to $\sim$\SI{2.6}{MeV}, it will be assumed that the ToF signal is produced only by protons. 
The mean proton spectrum and standard deviation retrieved from the ToF signals are shown in Fig.~\ref{Fig:Spectra}(b).

In order to quantify the stability and reproducibility of the ion source, the proton cut-off energy has been extracted from the reconstructed proton spectra, shown in Fig.~\ref{Fig:Spectra}(c), where the individual markers represent the cut-off energy for each individual irradiation, and the straight line depicts the 30-point moving average of the data. As it can be seen, the points are distributed around a mean cut-off energy of \SI{0.82}{MeV}, with a dispersion given by the standard deviation of \SI{0.12}{MeV}, representing a $\sim$15\% of relative dispersion. Albeit these values support the high reproducibility and stability of the long-term operation of the target wheel and proton source, the dispersion is larger than what would be expected considering the measured reproducibility of the target positioning on the focal plane. However, it should be noted that the different parameters of the laser, such as energy, pointing, and spatial phase, were not stabilised, and the fluctuations in the ion spectra and cut-off energy were significantly affected by shot-to-shot variations of the laser itself. Furthermore, the continuous irradiation of the laser results in the heating of the different optical elements along the beamline, which has been shown to rapidly affect the alignment and behaviour of the laser, in turn degrading the stability of the proton beam~\cite{alejo2022characterisation}. We conclude, therefore, that the variability of the proton spectrum is dominated by variations of the laser beam rather than by the operation and alignment of the target, and can be further reduced with suitable pulse stabilisation techniques.

\section{Conclusions}
Here we have presented a multi-shot target assembly for laser-plasma ion acceleration compatible with multi-Hertz operation. The assembly consists of a three-dimensional motorised rig, with one rotational and two linear stages that guarantee the shot-to-shot replenishing and positioning at laser focus of the target material, and a wheel target holder. The rotating wheel is capable of hosting $>$5000 targets and is designed to operate continuously at rates of up to \SI{10}{Hz}.

An automatic procedure for the alignment of the target surface with respect to the laser focal plane for each impact position has been introduced. This procedure is based on a three-dimensional pre-map of the desired shooting positions obtained prior to irradiation, with an industrial optical sensor with micron-level precision. This map of positions, which can be retrieved in a few minutes, can be used to calculate the required correction to ensure the target placement by the software controlling the shot-to-shot movement of the three stages of the target assembly. Following this procedure, we have demonstrated that the individual targets can be positioned at the laser focus with an accuracy of $\sigma=\SI{3.5}{\micro\m}$, significantly lower than the $\sim\SI{10}{\micro\m}$ precision required due to the Rayleigh length of the focusing system. This solution represents a significant boost in both the number of shots, competitive with respect to other solutions such as MEMS technology or tape-drive systems, and the repetition rate at which it can operate, without requiring prolonged periods for the manual pre-characterisation of the target surface.

The target assembly has been succesfully used to accelerate ions using the high-power laser system at the L2A2 facility. The laser-driven proton beam has been characterised by means of a time-of-flight detector. A stable, continuous laser-driven proton source was demonstrated from the irradiation of $>$1000 consecutive targets, exhibiting a mean cut-off energy of \SI{0.82}{MeV} and relative dispersion of 15\%, mainly due to the shot-to-shot variations in the laser parameters.

\section{Acknowledgements}
This work was supported by the Spanish Ministerio de Ciencia, Innovaci\'on y Universidades under grants RTI2018-101578-B-C21, RTI2018-101578-B-C22, FPI predoctorals BES-2017-08917 and PRE2019-090730, and Unidad de Excelencia Mar\'ia de Maetzu under project MdM-2016-0692-17-2, the Xunta de Galicia grant GRC ED431C 2017/54 and a grant of the program Grupos de investigación consolidados (CIAICO/2022/008), and EDGJID/2021/204 financed by Generalitat Valenciana. Action co-financed by the European Union through the Programa Operativo del Fondo Europeo de Desarrollo Regional (FEDER) of the Comunitat Valenciana 2014-2020 (IDIFEDER/2021/002). This work was supported by `la Caixa' Foundation (ID 100010434) [fellowship code LCF/BQ/PI20/11760027], and grant RYC2021‐032654‐I funded by MCIN, AEI and by `European Union NextGenerationEU'.

\bibliographystyle{ieeetr}
\bibliography{bibliography_ions} 

\end{document}